\documentclass[acmsmall,screen]{acmart}

\AtBeginDocument{%
  }

\usepackage{algorithmic}
\usepackage{graphicx}
\usepackage{textcomp}
\usepackage{xcolor}
\usepackage[most]{tcolorbox}
\usepackage{colortbl}
\usepackage{multirow}
\usepackage{adjustbox}
\usepackage{array}
\usepackage{longtable}
\usepackage{geometry}
\usepackage{xspace}
\usepackage[shortlabels]{enumitem}
\usepackage{booktabs}    %
\usepackage{tabularx} 
\usepackage[]{mdframed}
\usepackage{tcolorbox}
\usepackage{adjustbox}
\usepackage{pdflscape}
\usepackage{multirow}
\usepackage{multirow}
\usepackage{caption}
\usepackage{cleveref}
\crefformat{section}{\S#2#1#3} %
\crefformat{subsection}{\S#2#1#3}
\crefformat{subsubsection}{\S#2#1#3}

\usepackage{listings}
\usepackage{caption}

\definecolor{codegreen}{rgb}{0,0.6,0}
\definecolor{codegray}{rgb}{0.5,0.5,0.5}
\definecolor{codepurple}{rgb}{0.58,0,0.82}
\definecolor{backcolour}{rgb}{0.95,0.95,0.92}

\lstdefinestyle{mystyle}{
    backgroundcolor=\color{backcolour},
    commentstyle=\color{codegreen},
    keywordstyle=\color{magenta},
    numberstyle=\tiny\color{codegray},
    stringstyle=\color{codepurple},
    basicstyle=\ttfamily\footnotesize,
    breaklines=true,
    numbers=left,
    numbersep=7.5pt,
    tabsize=2,
    xleftmargin=0em,
    xrightmargin=0em,
    abovecaptionskip=7.5pt
}

\usepackage[ruled,vlined,linesnumbered]{algorithm2e}
\usepackage{amsmath}
\SetKw{Continue}{continue}

\newcommand*\numblack[1]{%
  \tikz[baseline=(char.base)]{
    \node[shape=circle,draw=black,fill=black, text=white,inner sep=1pt] (char) {#1};}}

\newcommand{\ourapproach}{\textsc{CovAgent}\xspace}
\newcommand{\nametool}{\textsc{CovAgent}\xspace}

\usepackage{soul}        %
\usepackage[colorinlistoftodos,prependcaption,textsize=tiny]{todonotes}

\setcopyright{acmlicensed}

\begin{document}

\title{\ourapproach: Overcoming the 30\% Curse of Mobile Application Coverage with Agentic AI and Dynamic Instrumentation}

\author{Wei Minn}
\email{wei.minn.2023@phdcs.smu.edu.sg}
\orcid{0000-0002-3191-9795}
\affiliation{%
  \institution{Singapore Management University}
  \country{Singapore}
}

\author{Biniam Fisseha Demissie}
\email{biniam.demissie@tii.ae}
\orcid{0000-0002-5369-5235}
\affiliation{%
  \institution{Technology Innovation Institute}
  \country{UAE}
}

\author{Yan Naing Tun}
\email{yannaingtun@smu.edu.sg}
\orcid{0009-0009-2899-4637}
\affiliation{%
  \institution{Singapore Management University}
  \country{Singapore}
}

\author{Jiakun Liu}
\authornote{Corresponding Author}
\email{jiakunliu@hit.edu.cn}
\orcid{0000-0002-7273-6709}
\affiliation{%
  \institution{Harbin Institute of Technology}
  \city{Harbin}
  \country{China}
}

\author{Mariano Ceccato}
\email{mariano.ceccato@univr.it}
\orcid{0000-0001-7325-0316}
\affiliation{%
  \institution{University of Verona}
  \city{Verona}
  \country{Italy}
}

\author{Lwin Khin Shar}
\email{lkshar@smu.edu.sg}
\orcid{0000-0001-5130-0407}
\affiliation{%
  \institution{Singapore Management University}
  \country{Singapore}
}

\author{David Lo}
\email{davidlo@smu.edu.sg}
\orcid{0000-0002-4367-7201}
\affiliation{%
  \institution{Singapore Management University}
  \country{Singapore}
}

\begin{abstract}
Automated GUI testing is crucial for ensuring the quality and reliability of Android apps. However, the efficacy of existing UI testing techniques is often limited, especially in terms of coverage. Recent studies, including the state-of-the-art, struggle to achieve more than 30\% activity coverage in real-world apps. This limited coverage can be attributed to a combination of factors such as failing to generate complex user inputs, unsatisfied activation conditions regarding device configurations and external resources, and hard-to-reach code paths that are not easily accessible through the GUI.
To overcome these limitations, we propose \nametool, a novel agentic AI-powered approach to enhance Android app UI testing. 
Firstly, our fuzzer-agnostic framework runs a standard and widely adopted GUI fuzzer such as APE and Fastbot to identify unreachable activities. 
Then, those unreachable activities are analyzed by an AI agent that inspects the app's decompiled smali code and component transition graph, and reasons about unsatisfied activation conditions within the app code logic that prevent access to the activities. 
Based on this analysis, another agent generates dynamic instrumentation scripts that are injected into the running app instance to satisfy activation conditions
required for successful transitions to the target activity.
In the evaluation, we show that augmenting existing fuzzing approaches with our framework achieves a significant improvement in test coverage. 
Compared to the state-of-the-art, LLMDroid, and other baselines such as Fastbot and APE, \nametool achieves 101.1\%, 116.3\% and 179.7\% higher activity coverage, respectively. \nametool also outperforms all the baselines in other metrics such as class, method, and line coverage. 
We also conduct investigations into components within \nametool to reveal further insights regarding the efficacy of Agentic AI in the field of automated app testing such as the agentic activation condition inference accuracy, and agentic activity-launching success rate.
To foster further research and development in this area, we open-source \nametool at~\cite{replicationpackage}.

\end{abstract}



\maketitle

\section{Introduction}
\label{sec:intro}

Automated testing of Android apps is an active area of research nowadays.
Android apps are event-driven programs with complex GUIs, and manual testing is labor-intensive and may overlook many corner cases.
To automatically test Android apps, generating meaningful event sequences (e.g., add items to the shopping cart before checking out) and producing semantically meaningful inputs (e.g., math operators for a calculator app) are necessary \cite{choudhary2015automated,yu2024practical}.

To fill the gap, researchers proposed a series of approaches to explore the app's GUI components (e.g., activities, fragments, dialogs) randomly (e.g., Monkey~\cite{monkey}), systematically (e.g., A3E~\cite{azim2013targeted}, Stoat~\cite{su2017guided}, Sapienz~\cite{mao2016sapienz}, APE~\cite{ape}), and intelligently (e.g., Humanoid~\cite{li2019humanoid}).
In addition, some approaches leverage program analysis to understand the app's code and generate test inputs accordingly ~\cite{ape,mahmood2014evodroid}.
Recently, with the advance of large language models (LLMs), some approaches leverage LLMs to understand the app's GUI and code and generate test inputs accordingly (e.g., DroidAgent~\cite{droidagent}, GPTDroid~\cite{liu2024make}, LLMDroid~\cite{llmdroid}).

Yet, challenges remain in terms of code coverage as studies show that even the state-of-the-art tools struggle when testing industrial-scale apps \cite{akinotcho2025curse, llmdroid}. Specifically, Akinotcho et al.~\cite{akinotcho2025curse} analyzed 103 industrial-scale apps and reported 10 new reasons why current testing approaches ``hit a ceiling'' of 30\% coverage.
Specifically, the launch of these activities
(1) needs the input from a third-party server, (2) needs to be accessed from the alternate entry (rather than starting from the main activity), (3) needs external resources (e.g., another equipments or additional information), (4) is disabled for a version or user, (5) is triggered by errors, (6) needs specific hardware/software, (7) needs certain environment (e.g., location), (8) needs a certain usage pattern (e.g., the number of times the users switch the channel), (9) requires the launch of the caller activity, and (10) is impossible because no entry point triggering the activity.

Theoretically, if we could design a tool to solve these constraints automatically, we could break the 30\% coverage curse.
We take Samsung Smart Switch app's \texttt{IosOtgContentsListActivity} activity as an example, which is not reachable by existing tools and is used to transfer app data between the device and an SD card.
To launch this activity, following the suggestion provided by Akinotcho et al.~\cite{akinotcho2025curse}, we need to:

\vspace{.25em}
\noindent \textbf{C1) Identify the activation conditions (e.g., guard conditions, and launch sites) of the unreachable activities.}
We observe that the \texttt{IosOtgContentsListActivity} contains a guard condition that checks for connection to SD card and terminates if there is no connection.
Listing~\ref{lstMoti} depicts decompiled Smali assembly code of the \texttt{onCreate} lifecycle hook inside \texttt{ContentsListBaseActivity} which is the parent class of \texttt{IosOtgContentsListActivity}. Line~14 calls \texttt{ShowNeedSdCardPopup} sub-procedure and line~15 stores its return value into \texttt{v0}. Afterwards, the conditional statement line~18 is the guard condition that checks if the return value stored in \texttt{v0} is \texttt{true} or \texttt{False}.
If \texttt{v0} is \texttt{true} from \texttt{ShowNeedSdCardPopup}, the \texttt{onCreate} procedure terminates immediately by calling the \texttt{return} statement in line~11, and the activity will not be successfully launched.
To identify this kind of code-level activation conditions, the latest work by Akinotcho et al.~\cite{akinotcho2025curse} involves decompiling the app's APK file using JADX decompiler~\cite{jadx} for the authors to manually inspect the decompiled code of the app, and identify guard conditions.
Not only this is time-consuming and labour-intensive, it also requires the analysts to possess thorough knowledge of the app's features and user scenarios, and how they could be represented inside the decompiled code.

    \begin{center}
\begin{minipage}{.8\linewidth}
        \begin{lstlisting}[caption={\textbf{\texttt{onCreate} lifecycle hook of \texttt{ContentsListBaseActivity}} that is defined inside \texttt{ContentsListBaseActivity.smali} and is the parent class of \texttt{IosOtgContentsListActivity} defined inside \texttt{IosOtgContentsListActivity.smali}}, label=lstMoti]
    // Inside ContentsListBaseActivity.smali
    .class public abstract ContentsListBaseActivity;
    .super ActivityBase;
    .source "ContentsListBaseActivity.java"
    ...
    .method protected onCreate(Landroid/os/Bundle;)V
        ...
        // logic for terminating
        :cond_0
        :goto_0
        return-void
        ...
        // calling subprocedure
        invoke-virtual {p0}, ContentsListBaseActivity;->ShowNeedSdCardPopup()Z
        move-result v0
        ...
        // guard condition to check the return value
        if-nez v0, :cond_0
        ...
    ...
    
    // Inside IosOtgContentsListActivity.smali
    .class public IosOtgContentsListActivity;
    .super ContentsListBaseActivity;
    .source "IosOtgContentsListActivity.java"
    ...
    \end{lstlisting}
\end{minipage}
    \end{center}

\begin{figure}
    \includegraphics[width=\textwidth, trim=0.5cm 0.5cm 0.5cm 0.3cm]{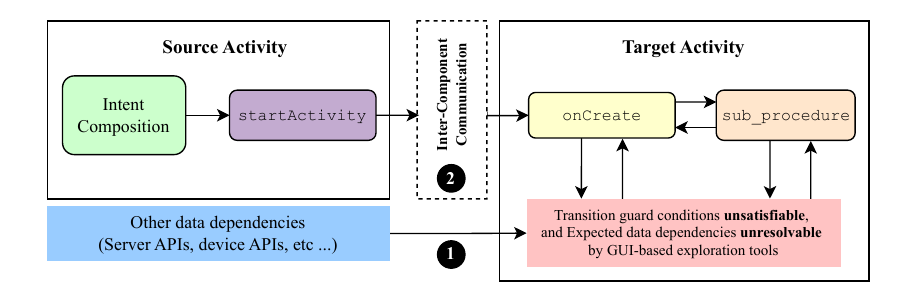}
    \caption{\textbf{Data dependencies of an app activity.} While an app activity can require data from received ICC messages from source components such as caller activities, it can also require data from non-ICC sources such as APIs for communicating with servers and devices.}
    \label{fig:motivation}
\end{figure}

\noindent \textbf{C2) Develop techniques for generating data values needed for successful activation of an activity.}
We observe that the \texttt{onCreate} of \texttt{IosOtgContentsListActivity} requires \texttt{ShowNeedSdCard} \texttt{Popup} sub-procedure to return \texttt{false} for the activity to successfully initialize.
Figure~\ref{fig:motivation} depicts the logical flow where data for successful launch depends on APIs that are not related to Intents (\numblack{1}).
Existing works for activity launching such as~\cite{scenedroid, fax, dalt, stimulation, columbus, car} use static analysis to focus on generating only Intent-based test cases (\numblack{2}).
However, they do not cover data dependencies on the external resource-related APIs (in the case of \texttt{IosOtgContentsListActivity}), or any of the 9 other reasons presented by Akinotcho et al.~\cite{akinotcho2025curse}.

\vspace{.5em}
\noindent \textbf{C3) Develop appropriate strategies (e.g., ADB commands or generating code-level activation) to resolve the activation conditions automatically.}
We observe that \texttt{IosOtgContentsList} \texttt{Activity} requires the device's SD card connection state to be in a particular way; existing approaches for launching activities~\cite{scenedroid, columbus, car, fax, dalt, brahmastra, stimulation} use the Android Debug Bridge (ADB) which does not support the manipulation of the app's perception of device state or external resources to bypass the guard conditions at the start of the activity. 
Moreover, they face more challenges in terms of limited support by the ADB tool in composing non-trivial Intent messages, such as Intent Bundles, and sending non-primitive objects through the command line.
Existing literature~\cite{akinotcho2025curse} also reports instability issues owing to the invasive nature of existing approaches~\cite{car, stimulation, columbus} (e.g. instrumenting AOSP class loader source code, or modifying \texttt{DEX} bytecode). 

\vspace{.5em}
To automatically launch the unreachable activities, we propose \nametool, a novel approach that leverages dynamic instrumentation scripts (using Frida framework~\cite{fridatool}), static and dynamic program analysis, and LLM to address the above challenges.
Note that our goal is not to develop yet another GUI testing approach, but rather to create a novel accompanying framework that can satisfy the conditions needed to trigger activities unreachable by existing GUI fuzzing approaches.

To address \textbf{C1}, \nametool first extracts the Component Transition Graph (shortened as CTG) and decompiles the app to obtain relevant information (e.g., control and data dependencies) about the ICCs and other APIs invocations inside the unreached activities.
The advancement of LLMs and LLM-powered agents has shown great potential in various areas of software engineering~\cite{llm4se, agent4se}. 
Moreover, Model-Context Protocol (MCP) framework standardizes the interface between an LLM-based agent and its tools and resources \cite{mcpprotocol}.
In our case, we employ MCP framework to give an LLM-powered agent access to external tools such as program analysis tools to reason about the app's code such as the extracted CTG and decompiled code.
\nametool then uses a Chain-of-Thought (CoT) prompting technique to guide the agent to investigate about the possible reasons for the unreachability of each activity by forward and backward analysis with the help of MCP tools, and provide activation conditions that needs to be satisfied for a successful launch of the activity.
To address \textbf{C2}, \nametool instructs the agent in the same prompt to use the information in the activation condition that was gathered from forward and backwards analysis to generate data values required by the activity for a successful launch. Looking at the names and usages of invoked APIs, the agent ascertains the value that can satisfy the guard condition for a successful initialization of the activity.

Finally, to address \textbf{C3}, we design \nametool to use Frida instrumentation scripts to invoke the activities directly at the code level, thus bypassing the need for ADB commands. Frida is an industrial strength dynamic instrumentation tool used by prior works for testing and discovering vulnerabilities in Android apps \cite{otphunter, appprotect} without having to modify the AOSP source code. 
The instrumentation scripts hook into the app's process and modify the implementation of the app's methods inside memory to manipulate the app's perception of device state or external resources.
To achieve this, \nametool employs another agent to generate the instrumentation scripts using the activation conditions inferred by the previous agent. This script generation agent also has access to the parts of app code that was inspected by the previous agent that is useful for generating functional instrumentation scripts. To further improve the robustness and stability of the generated instrumentation scripts, we validate those scripts inside the Android emulator, and provide the feedback to the agent in terms of exception messages to refine the instrumentation scripts.

To evaluate the performance of our approach, we aim to answer the following research questions:

\vspace{.25em}
    \noindent \textbf{RQ1 (Breaking the Ceiling):} Can \nametool break the 30\% activity coverage ceiling reported by Akinotcho et al. ~\cite{akinotcho2025curse}?
 
This RQ aims to discover if the instrumentation scripts generated by \nametool can enhance the activity coverage by an existing UI fuzzing tool, more specifically, APE which is the same tool used by Akinotcho et al. to empirically determine the 30\% activity coverage ceiling. 
Results show that \nametool-APE, i.e., APE augmented with \nametool, achieves excellent results in activity coverage compared to APE alone, reaching an average of 49.5\% activity coverage versus 17.7\% by APE. 

\vspace{.25em}
    \noindent \textbf{RQ2 (\nametool vs. State-of-the-art):} How does \nametool compare to the state-of-the art in LLM-enhanced Android fuzzing, namely, LLMDroid in activity coverage?

This RQ aims to directly compare with the latest state-of-the-art for LLM-driven tool that also enhances an existing tool to improve coverage.
We choose to compare against LLMDroid-Fastbot specifically as it is the most performant of all other available tool combinations with LLMDroid. 
Results show that \nametool-Fastbot significantly improves test coverage over the state-of-the-art tool, with \nametool-Fastbot achieving average activity coverage of 34.6\% vs 17.2\% by LLMDroid-Fastbot.

\vspace{.25em}
    \noindent \textbf{RQ3 (Activity Launching Success):} How does \nametool perform in terms of activity launching success rate compared to the state-of-the-art Scenedroid?

This RQ investigates the activity launching effectiveness of the Frida instrumentation scripts generated by \nametool. We compare \nametool against Scenedroid~\cite{scenedroid} which is the state-of-the-art for intent-based activity launching that was discussed in \textbf{C2} and \textbf{C3}.
The experimental results show that \nametool's activity launching capability through its dynamic edge instrumentation agent achieves a 54.8\% activity launch success rate, vastly outpacing Scenedroid's 15.8\%. %

\vspace{.25em}
    \noindent \textbf{RQ4 (Activity Activation Condition Inference):} How well can \nametool's agentic activity activation condition inference detect the ground truths provided by Akinotcho et al.?

This RQ aims to validate whether an automated, agent-based approach can accurately identify activity activation conditions, and can serve as a reliable alternative to labor-intensive manual investigation.
Since this work is the first in automated activation condition extraction, we select a random baseline as a sanity check.
This identification of activation condition is quantified by recall score measuring how many of the unreachability reason(s) manually labelled by Akinotcho et al. are detected.
Results show that \nametool significantly outperforms the random baseline in the detection of unreachability reasons.

\vspace{.25em}
\noindent \textbf{Contributions.} This paper makes the following contributions:
\begin{itemize}[nosep,leftmargin=1em]
    \item \textbf{Exploration Framework.} A novel Android UI fuzzing approach that addresses the challenges of dependency conditions on server, external resources, environment, and device and other conditions, using a hybrid static-dynamic analysis coordinated with agentic AI solution for automation and test efficiency.
    \item \textbf{Experiments.} An empirical study that demonstrates 1) \nametool approach's coverage improvement over existing GUI fuzzing tools such as APE and Fastbot, and also the state-of-the-art in LLM-enhanced exploration, LLMDroid, 2) activity launching success rate improvement over state-of-the-art activity launching approach, Scenedroid, and 3) superior accuracy of agentic activation condition inference over random baseline.
    \item \textbf{Replication package.} We open-source our tool, \nametool, along with all the datasets, and detailed evaluation results at~\cite{replicationpackage}.
\end{itemize}

\section{Related Work}
\label{sec:related}

\subsection{Automated Android GUI Testing}

Automated GUI testing of Android apps is a well established research area. 
Traditionally, those approaches mainly focused on two primary challenges: maximizing GUI exploration and generating meaningful, human-like event sequences. 
To address exploration challenge, foundational approaches employed random input generation strategies such as Monkey~\cite{monkey} and DroidBot~\cite{li2017droidbot}, model-based techniques that build some form of representation of the app's GUI~\cite{guiripper-ase12,ape,su2017guided,lv2022fastbot2,choi2013guided,dong2020time,mirzaei2016reducing,yang2018static, azim2013targeted, jue2020combodroid} or search-based algorithms that formulate the exploration maximization as an optimization problem~\cite{mao2016sapienz,mahmood2014evodroid}. 
Approaches such as Sapienz~\cite{mao2016sapienz}, APE~\cite{ape}, Stoat~\cite{su2017guided}, ComboDroid~\cite{jue2020combodroid}, and Fastbot2~\cite{lv2022fastbot2} also focused on generating meaningful input sequences to uncover deeper functionalities. 
For example, Sapienz analyzed the effects of atomic events, e.g., click, pinch zoom, system events, and motif patterns that combine different atomic events. 
APE built a static GUI model and updates the model during testing to generate test cases based on the model to explore various sequences of events. 
ComboDroid starts from use cases representing distinct app functionalities, and systematically enumerated all possible combination of these use cases to create meaningful test input sequences (scenarios). 
While these approaches attempt to maximize GUI exploration by generating meaningful test input sequences using different techniques, \textbf{\nametool complements them by addressing the challenges faced by these approaches exemplified in Section~\ref{sec:intro} to enhance their activity coverage.}

\subsection{LLM-enhanced Android Testing}

More recently, the advent of Large Language Models (LLMs) has introduced a new opportunity for enhancing semantic understanding in testing and addressing some of the limitation of traditional approaches. Approaches such as QTypist~\cite{liu2023fill} leverage an LLM fine-tuned with labeled app UI screenshots to generate semantically valid GUI inputs. GPTDroid~\cite{liu2024make} formulates testing as a question-answering (Q\&A) task to generate action sequences for current functionality under test. 
DroidAgent~\cite{droidagent} performs intent-driven testing by leveraging multiple LLM instances. It prompts a LLM instance to first set testing goals and plan testing tasks, then utilizes another instance to execute these tasks, and conduct a reflection phase to make informed testing decisions about long term plans to achieve the testing goals. 
LLMDroid~\cite{llmdroid} is the state-of-the-art approach that uses a GUI fuzzer to autonomously explore app UI and employs an LLM to summarize the explored pages. It then uses the LLM to identify new UIs and functionalities that should be further explored and guides the fuzzer with this information to improve the coverage.
Similar to ours, these LLM-driven approaches such as QTypist~\cite{liu2023fill}, DroidAgent~\cite{droidagent}, LLMDroid~\cite{llmdroid} were also designed to be integrated with existing fuzzing tools like Monkey~\cite{monkey} and APE~\cite{ape}. \textbf{While many of these LLM-driven tools require modification of their accompanying GUI fuzzer, our approach is fuzzer-agnostic and integrates with any existing fuzzer out of the box.}
Thus, it is important to note that \ourapproach is not a competing approach with existing GUI fuzzers approaches (LLM-enhanced or otherwise), but, rather, an independent and complementary tool for dealing with those specific challenges, which can be seamlessly integrated in an exploration set  that already employs existing GUI-fuzzers such as APE and Fastbot.

\section{Approach}
\label{sec:approach}

\begin{figure*}[ht]
    \includegraphics[width=\textwidth, trim=0.75cm 0.75cm 0.75cm 0.75cm]{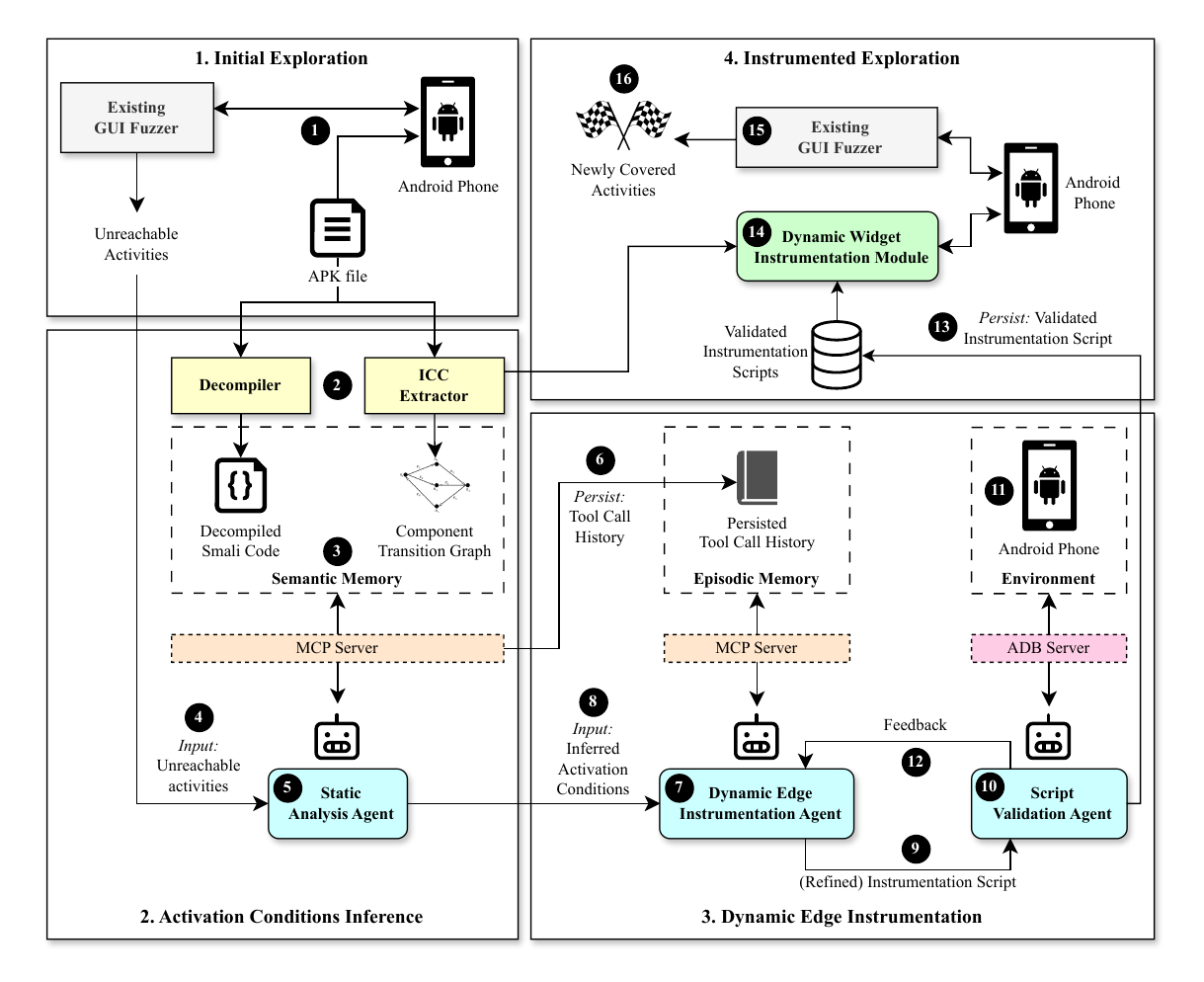}
    \caption{\textbf{Overview of \nametool.} After initial exploration using an existing GUI fuzzer, the static analysis agent explores the static features of the app to infer their activation conditions of unreached activities. The dynamic edge instrumentation agent uses these inferred activation conditions along with static features related to unreached activities to generate Frida instrumentation script. The script validation agent validates the generated instrumentation scripts in an emulated Android device. Finally, \nametool loads the validated scripts for all the target activities of the app into the app's process, and runs the GUI fuzzer concurrently to reach the target activities.}
    \label{fig:overview}
\end{figure*}

Figure~\ref{fig:overview} shows the workflow of our framework, \ourapproach.
Firstly, it runs an off-the-shelf GUI fuzzer to collect the reachability details of the activities in an app.
Then, given the unreachable activities and the static code features (e.g., component transition graph and decompiled code), \ourapproach uses the static analysis agent to autonomously infer (1) the reasons for the unreachability, as well as (2) the guideline for satisfying the activation conditions required to successfully launch the activities.
After that, with these two pieces of information, \ourapproach uses the dynamic instrumentation agent to generate a draft binary instrumentation script to dynamically hook into the app to satisfy the required activation conditions.
To validate the generated instrumentation script, \ourapproach uses the script validation agent to run the script on an Android emulator to check if the script can successfully trigger the transition to the target activity. 
If the transition fails, \ourapproach will use the feedback (e.g., error messages) from the script validation agent as an additional input of the dynamic edge instrumentation agent to refine the instrumentation script.
Finally, instrumentation scripts that can successfully transition to target activities are collated to instrument relevant activities in the app. The GUI fuzzer used in the initial step runs concurrently to explore the previously unreachable activities.

\subsection{Initial Exploration}
\label{sub:initial}

Our approach starts with collecting the unreachable activities.
\nametool runs an existing GUI fuzzer, such as APE, Stoat, or FastBot (Step \numblack{1} of Figure~\ref{fig:overview}).
This is because \ourapproach is fuzzer-agnostic, and hence, this initial exploration phase is loosely coupled to the rest of the framework.
We then derive unreachable activities (\emph{target activities} from hereon) from the set of all activities declared by the app (i.e., the set of activities extracted from the AndroidManifest file) minus those visited by the fuzzer.

\subsection{Activation Conditions Extraction}
\label{sub:activationextraction}

After collecting the unreachable activities, \ourapproach infers the reasons for the unreachability.

\subsubsection{Extracting static code features with static analyzers}
\label{sub:features}
To understand how to theoretically access the current unreachable activities, \ourapproach extracts the Component Transition Graph (CTG) of the app with the help of a static analyzer, i.e., ICCBot \cite{yan2022iccbot}.
CTG is a graph of the caller-callee relationship between components such as activities and fragments (Step \numblack{2}).
We use ICCBot as it is the state-of-the-art ICC resolution tool%
~\cite{yan2022comprehensive}.
As discussed in Section~\ref{sec:intro}, data dependencies on external resources and other APIs must also be accounted for when analyzing the app's code. A straightforward way to reveal those dependencies is by inspecting the body implementations of methods inside the target activity.
Thus, \ourapproach decompiles the app to obtain its Smali code and parses the code to obtain control- and data-flow information inside those methods.
We use Smali code because its code is readily extractable from any given APK file due to it being the "Assembly" for the Dalvik Virtual Machine of Android Runtime.
Finally, combining the CTG and the control- and data-flow information, we obtain the context to launch the activities.

\subsubsection{Accessing semantic memory of static code features} 
\label{sub:semantic}

The LLM's default limited knowledge of the app's procedures could lead to hallucinations and imprecise analysis of activation conditions, i.e., knowledge boundary problem \cite{knowledgeboundary}.
To solve this, prior studies in machine learning explored retrieval-based systems to provide LLMs with appropriate context~\cite{rag}. 
Semantic memory~\cite{semantic}, an application of retrieval systems in agentic AI, is an external data store that would be queried by LLM and would be the reasoning engine of our static analysis agent in the Section~\ref{subsub:static}.
Specifically, the static analysis agent will access the semantic memory (Step \numblack{3}) containing the static code features obtained from Section~\ref{sub:features} to aid its inference of the activation condition. Access to the semantic memory is enabled through the use of the MCP server, and the MCP tools available are described in Table~\ref{tab:mcp-tools-static}.

\subsubsection{Static Analysis Agent}
\label{subsub:static}

In order to infer activation conditions for a given target activity (Step~\numblack{4}), \nametool uses the static analysis agent (Step~\numblack{5}), powered by LLM, to autonomously investigate the application's code by accessing static code features stored inside its semantic memory.
This inference of the activation conditions involves forward analysis and backward analysis,
and the agent is instructed to \emph{think} in chain-of-thought (CoT)~\cite{wei2022chain} fashion to identify the dependencies and preconditions for a successful launching of the target activity. 
This CoT instruction depicted in Table~\ref{tab:prompt-designs-static} Row $2$ prompts the LLM to produce intermediate reasoning steps that allow it to solve complex tasks by breaking them into manageable tasks. %
Finally \nametool generates instrumentation guidelines that will be used in Section \ref{sub:edgeinstrumentationagent}:

\vspace{.25em}
\noindent
\textbf{Forward Analysis.}
To infer the activation conditions that are related to the initialization process of the target activity, 
we instruct the agent to start investigating from \texttt{onStart}, \texttt{onCreate}, and \texttt{onResume} lifecycle hooks of the target activity to ensure the inspection of their potentially exception-raising sub-procedures.
These lifecycle hooks typically contains the initialization logic which reveals key data- and control-flow dependencies from expected ICC messages (via \texttt{getIntent()}), and other APIs and global variables in the app instance.
This approach is inspired by Fax~\cite{fax} that utilizes static analysis and constraint-solving to compose ICC messages required to launch an activity.
\nametool extends that idea to support additional data dependency scenarios such as the invocation of external resources and server APIs that cause app-crashing exceptions inside \texttt{onCreate}, \texttt{onStart}, and \texttt{onResume}, or their sub-procedures when not handled safely during app testing. 
\nametool also instructs the agent to recursively inspect the sub-procedures that are invoked by these lifecycle hooks to collect a more complete and deeper picture of control- and data-dependencies on the sub-procedures. 
This recursive investigation terminates when the static analysis agent has decided that it has collected enough information about the activation conditions of the target activity. 
This decision is done autonomously without any explicit rule given inside the instruction prompt.

\begin{table}[t]
\caption{Prompt Engineering for \textsc{Static Analysis Agent}}
\vspace{-1mm}
\label{tab:prompt-designs-static}
\renewcommand{\arraystretch}{1.4}
\setlength{\tabcolsep}{5pt}
\adjustbox{width=\textwidth,center}{%
\footnotesize %
\begin{tabular}{|c|l|p{9cm}|}
\hline
\textbf{ID} & \textbf{Prompt Type} & \textbf{Instantiation} \\
\hline
1 & Package Info & 
You are a Smali language (Dalvik assembly) expert \dots. Your task is to analyze the dependencies and preconditions for successfully launching a given activity in the \texttt{package\_name} app that was not covered during the previous automated GUI-driven app exploration.
\\
\hline
2 & CoT Instruction & 

\textbf{Step 1: Forward Analysis}

Inspect the body of onCreate, onStart or onResume lifecycle hooks of the target activity

- Inspect control-flow of the lifecycle hook body implementations \dots

\dots

\textbf{Step 2: Backward Analysis}

Inspect the body of source methods that call the `startActivity` to launch the target activity.

- If launching source method is available, analyze variables declared and \dots

\dots

\textbf{Step 3: Launching the target activity}

Derive the desired logical flow of statements achievable via Frida instrumentation to directly launch the target activity successfully using `startActivity`.

\dots
\\
\hline
3 & Prompt Generation Rule & Package Info + Chain-of-Thought + Target Activity Name\\
\hline
4 & Example Output & 
Based on my analysis, let me now summarize the findings and provide the dependencies and preconditions for successfully launching the \texttt{activity\_name}.

\textbf{\#\# Detailed Analysis Results}

\textbf{\#\#\# Forward Analysis}

\dots

\textbf{\#\#\# Backward Analysis}

\dots
\\
\hline
\end{tabular}
}
\vspace{2mm}
\small 
\end{table}

\vspace{.25em}
\noindent
\textbf{Backward Analysis. }
While the forward analysis step helps in activation conditions of the target activity such as the \emph{unpacking} of ICC messages (Intents), this step is concerned about the \emph{construction} of ICC messages that are used to launch the target activity. 
As such, the agent is instructed to perform backward analysis to inspect the call sites that launch the target activity (i.e., \texttt{startActivity} and \texttt{startActivityForResult} APIs.) .

\vspace{.25em}
\noindent
\textbf{Generating guidelines for a successful launch. }
With these two analysis techniques, the agent is equipped with the necessary information on how to satisfy the activation conditions of the target activity such as ICC messages (Intent extras, action strings, categories, and data URIs) and values from other APIs expected by \texttt{onCreate}, \texttt{onStart}, and \texttt{onResume} and their sub-procedures.
Finally, the agent produces a description on how and which procedures can be dynamically instrumented (as shown in Table~\ref{tab:prompt-designs-static} Row $4$), which will be used by the Dynamic Instrumentation Agent in Section~\ref{sub:edgeinstrumentationagent}.

\begin{table}[t]
\centering
\caption{MCP Tools for Accessing Semantic Memory}
\vspace{-1mm}
\label{tab:mcp-tools-static}
\renewcommand{\arraystretch}{1.4} %
\adjustbox{width=\textwidth,center}{%
\footnotesize %
\begin{tabular}{|p{1.7cm}|p{2.7cm}|p{2.6cm}|p{6.0cm}|}
\hline
\textbf{Static}\newline \textbf{Feature} & \textbf{MCP Tool} & \textbf{Parameters} & \textbf{Description} \\
\hline
\multirow{7}{*}[0pt]{Method body} 
& \texttt{get\_activities} & - & Returns the list of all declared activities within the application. \\
\cline{2-4}
& \texttt{check\_activity\_}\newline\texttt{exists} & 1. \texttt{class\_name}: string & Checks whether a given activity is defined in the app. \\
\cline{2-4}
& \texttt{check\_class\_exists} & 1. \texttt{class\_name}: string & Verifies the existence of any class within the decompiled Smali files. \\
\cline{2-4}
& \texttt{get\_methods\_} \newline \texttt{inside\_class} & 1. \texttt{class\_name}: string & Lists all methods defined within a specified class. \\
\cline{2-4}
& \texttt{get\_method\_body} & 1. \texttt{class\_name}: string, \newline2. \texttt{method\_sig}: string & Retrieves the full Smali code body for a given method. \\
\cline{2-4}
& \texttt{get\_methods\_invoked} & 1. \texttt{class\_name}: string, \newline2. \texttt{method\_sig}: string & Returns the list of methods invoked within a specified method body, enabling forward control-flow traversal. \\
\cline{2-4}
& \texttt{get\_caller\_methods} & 1. \texttt{class\_name}: string, \newline2. \texttt{method\_sig}: string & Identifies methods that call the specified method, supporting backward data- or control-flow traversal. \\
\cline{2-4}
& \texttt{get\_launching\_} \newline \texttt{activities\_and}\newline\texttt{\_methods} & 1. \texttt{target\_activity}:\newline string & Returns a list of activity-method pairs that invoke the target activity via startActivity.  \\
\hline
CTG & \texttt{get\_launching\_} \newline \texttt{activities\_and}\newline\texttt{\_methods} & 1. \texttt{target\_activity}: \newline string & Similar to its Smali counterpart, this function returns source activities and method signatures responsible for launching the target activity. \\
\hline
\end{tabular}
} %
\end{table}

\subsection{Dynamic Edge Instrumentation}
\label{sub:instrumetation}

Here, \nametool aims to generate valid Frida scripts that fulfill those activation conditions. Specifically, \nametool first uses an agent to generate the Frida script, and uses another agent to validate the generated scripts. The failure information from validation will be used as feedback to regenerate the Frida script.

\subsubsection{Dynamic Edge Instrumentation Agent}
\label{sub:edgeinstrumentationagent}

To launch the target activities successfully, we introduce a dynamic edge instrumentation agent that generates a Frida instrumentation script.
This frida instrumentation script will be loaded into a running app process inside an Android device to trigger a transition to the target activity.

The dynamic edge instrumentation agent is an LLM-driven agent that interprets activation conditions inferred in Section~\ref{sub:activationextraction} to generate the instrumentation script. Just like the static analysis agent discussed in Section~\ref{subsub:static}, the dynamic edge instrumentation agent's LLM also suffers from a knowledge boundary problem and risks producing invalid instrumentation scripts as a result of hallucination. Thus, it also needs to be given access to the static code features to aid its script generation process. To this end, the static analysis agent from the previous phase has already explored the relevant static code features that are related to the activation condition of the target activity. To access only the static code features related to the target activity and avoid spending extra effort and tokens by exploring the code base again, inspired by works in agentic AI \cite{multiagent} and reinforcement learning\cite{sutton_barto} that collect experiences from past actions, we limit the dynamic edge instrumentation agent to access only those static code features that have been accessed by the static analysis agent during the prior phase of activation condition inference, i.e., we introduce episodic memory. 
In our case, the episodic memory is a data store that contains MCP tool call results that were queried during the activation condition inference phase for \emph{the respective target activity only}. 
This is achieved by using the static analysis agent's logging features to persist its tool queries and results into a file (\numblack{6}), and is available to the dynamic instrumentation agent via a database query or an MCP tool call.

\begin{table}[t]
\caption{Prompt Engineering for \textsc{Dynamic Instrumentation Agent}}
\vspace{-1mm}
\label{tab:prompt-designs-dynamic}
\renewcommand{\arraystretch}{1.4}
\setlength{\tabcolsep}{5pt}
\adjustbox{width=\textwidth,center}{%
\footnotesize %
\begin{tabular}{|c|l|p{9cm}|}
\hline
\textbf{ID} & \textbf{Prompt Type} & \textbf{Instantiation} \\
\hline
1 & CoT Instruction & 

\textbf{Step 1: Read \texttt{state['phase']}, and 1) if the value of \texttt{state['phase']} is \texttt{Generate}, analyze extracted activation conditions and instrumentation strategy suggestions; 2) if the value of \texttt{state['phase']} is \texttt{Refine}, evaluate the most recent generated code in \texttt{state['current\_code']} \dots}

For both cases, if need to, you can call the tool \texttt{retrieve\_tool\_call\_result} to inspect the results \dots

- \texttt{get\_activities()}: Get the list of activities of the app.

- \texttt{check\_activity\_exists(activity\_name: str)}: Check if an activity exists inside the app.

\dots

\textbf{Step 2: Write down the logic of the instrumentation in pseudocode.}

\dots

\textbf{Step 3: Translate the pseudocode into a valid Frida instrumentation JavaScript code.}

\dots
\\
\hline
2 & Tool Call History & 
Tool Calls made by Static Analysis agent while analyzing dependencies of \texttt{activity\_name}:

1: name: \texttt{check\_activity\_exists}, args: \{

\hspace*{2em}'\texttt{activity\_name}': '\texttt{com.fsck.k9.activity.ChooseAccount}'

\}

2: name: \texttt{get\_methods\_inside\_class}, args: \{

\hspace*{2em}'\texttt{class\_name}': '\texttt{com.fsck.k9.activity.ChooseAccount}'

\}

\dots
\\
\hline

3 & Activation Conditions & 
Final response of Static Analysis agent after analyzing activation conditions to enable the coverage of \texttt{com.fsck.k9.activity.ChooseAccount} in the next round of automated GUI-based app exploration:

\dots
\\
\hline
3 & Prompt Generation Rule & CoT Instruction + Tool Call History + Activation Condition \\
\hline

4 & Example Output & 
\texttt{Java.perform(function() \{}

\hspace*{2em}\texttt{try \{}

\hspace*{4em}\texttt{console.log("[*] Instrumenting app");}

\hspace*{4em}\dots
\\
\hline
\end{tabular}
}
\vspace{2mm}
\small 
\end{table}

To generate the Frida instrumentation script for launching the target activity, we prompt the agent (\numblack{7}) with CoT~\cite{wei2022chain} to instruct the agent to incorporate the context throughout the script generation process as shown in Table~\ref{tab:prompt-designs-dynamic} Row 1. 
In Row 2, the persisted MCP tool call results (\numblack{6}) are given as context to the dynamic analysis agent
In Row 3, we include the instruction for generating a pseudo-code that describes the logic of the instrumentation script before generating the actual instrumentation script (\numblack{8}), and use it as a guideline to generate the actual Frida script. It has been demonstrated that prompting LLMs to generate pseudo-code before generating the actual code improves their code generation performance~\cite{huang2023agentcoder}.
Finally, in Row 4, the agent generates the pseudo-code of the instrumentation script and translates it into actual Frida instrumentation code (\numblack{9}) that is ready to be loaded into a running app process.

\subsubsection{Script Validation Agent}
\label{sub:validationagent}

\begin{figure*}
    \includegraphics[width=.75\textwidth, trim=0cm 0.25cm 0cm 0.7cm]{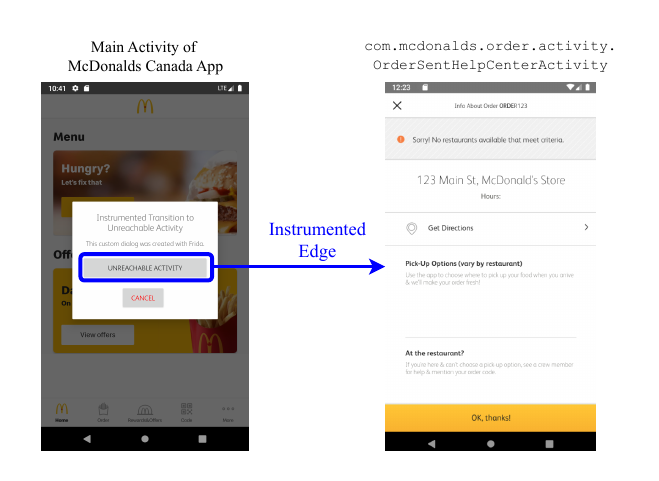}
    \vspace{-1em}
\captionsetup{justification=centering}
    \caption{Injected button widget (on the left) that invokes transition to the target activity (on the right) to validate the edge instrumentation script}
    \label{fig:scriptvalidation}
\end{figure*}

In order to validate the edge instrumentation script generated by the dynamic edge instrumentation agent, we introduce a script validation agent (\numblack{10}) that validates the effectiveness of the generated script by injecting it into the running app's process (\numblack{11}) to launch the target activity.
The core idea is: the script validation agent injects a pop-up button widget into the main activity of the app, whose \texttt{onClick} event listener invokes the script generated by the dynamic instrumentation agent that navigates to the target activity as shown in Figure~\ref{fig:scriptvalidation}.
Once the script validation agent clicks on the button, the target activity will be shown.\footnote{Note that in the Instrumented Exploration phase (Section~\ref{sub:instrumentedexploration}), we extend this idea to instrumenting for multiple target activities (instead of single target) for the existing GUI fuzzer to explore all the target activities during exploration.}
Note that this script validation agent is not LLM-driven and, in addition to script validation, also handles the communication with the device, such as booting, installing the app, and the Frida server.

\vspace{.5em}
\noindent
\textbf{Feedback.}
In order to aid the refinement of the instrumentation script by the dynamic edge instrumentation agent, the script validation agent provides feedback with details of the failed transition (\numblack{12}).
The feedback is composed using information from three points of observation inside the device:
\emph{1) Frida instrumentation exception:}
If there are runtime errors when loading the generated Frida script into the app process, the Frida server process inside the device throws an exception. 
This can happen if the script attempts to invoke non-existent methods or classes because of a dynamic instrumentation agent hallucinating during the script generation process and making imprecise instrumentation targets.
\emph{2) App crash:}
The instrumented app can encounter a runtime exception that typically leads to app crashes, thus preventing a complete launch of the target activity. 
This happens when the dynamic instrumentation agent does not have complete knowledge of what instrumentation to apply or generates an incomplete instrumentation script for the sake of brevity, thus missing required data or expected Intent messages. In this case, an exception trace can usually be found in the device log.
\emph{3) No transition:}
If the transition to the target activity does not happen, but there are no relevant exception traces inside the application log or the Frida server, a feedback message about the failed transition is still sent back to the dynamic instrumentation agent to refine the script.

To save time and resources, we set a maximum of five iterations for this feedback loop. If the target activity is launched successfully, the activity is marked as success, and the loop is terminated without sending any feedback back to the dynamic instrumentation agent. After five iterations, if the target activity cannot be launched, the loop terminates, and the activity is considered unreachable by \nametool. 

\subsection{Instrumented Exploration}
\label{sub:instrumentedexploration}

\begin{algorithm}[t]
\caption{Dynamic Widget Instrumentation Algorithm}
\label{alg:widgetinstrumentation}

\KwIn{Available instrumentations \textit{instrumentations}, Component Transition Graph \textit{ctg}, \newline Main activities \textit{mains}, Unreachable activities \textit{unreachables} }
\KwOut{Activity dialogs $activity\_dialogs$}

\SetKwProg{Fn}{Function}{}{}

\Fn{\textsc{FindDialogForTarget}}{
    $activity\_dialogs \leftarrow \{\}$
    
    $found\_reachable \leftarrow \emptyset$

    $non\_main\_reachables \leftarrow mains - unreachables$
    
    \For{$target\_activity \in instrumentations$}{

        \If{$target\_activity \notin unreachables$}{
            \Continue
        }

        \If{$ctg.\textnormal{getSources}(target\_activity) = \emptyset$}{
            \For{$main \in mains$}{
                $activity\_dialogs[main].\textnormal{add}(target\_activity)$
            }
            \Continue
        }

        \For{$source\_activity \in ctg.\textnormal{getSources}(target\_activity)$}{
            \If{$source\_activity \in non\_main\_reachables$}{
                $activity\_dialogs[source\_activity].\textnormal{add}(target\_activity)$
                
                $found\_reachable.\textnormal{add}(target\_activity)$
            }
        }

        \If{$target\_activity \notin found\_reachable$}{
            \For{$main \in mains$}{
                $activity\_dialogs[main].\textnormal{add}(target\_activity)$
            }
        }
    }
}
\end{algorithm}

To enhance the coverage of existing GUI fuzzers (e.g., APE and Fastbot), we load the validated instrumentation scripts into the respective app process inside the Android emulator. Concurrently, in a separate process, the fuzzer is run to use the loaded instrumentation scripts to visit the target activities that are initially unreachable to the tool in Section~\ref{sub:initial}.

\subsubsection{Dynamic Widget Instrumentation}
\label{sub:widget}
\begin{figure*}
    \includegraphics[width=.75\textwidth, trim=0cm 0.25cm 0cm 0.7cm]{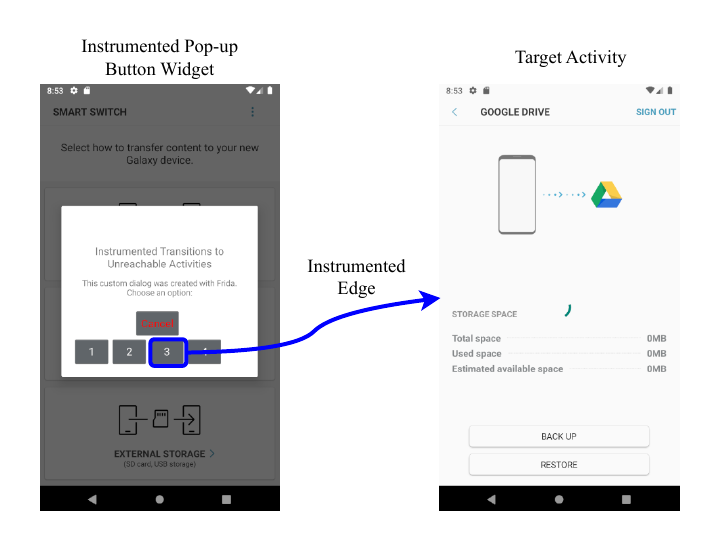}
    \vspace{-1em}
\captionsetup{justification=centering}
    \caption{Pop-up dialog that is injected into Samsung Smart Switch app by \nametool contains buttons that trigger instrumented edges}
    \label{fig:instrumentedexploration}
\end{figure*}

In order for the GUI fuzzer to cover all the target activities for which instrumentation scripts are validated (Step \numblack{13} of Figure~\ref{fig:overview}), \nametool uses Frida server inside the Android emulator to load all the validated instrumentation scripts at the same time.
To achieve this, we introduce a dynamic widget instrumentation module that extends the script validation agent's injection of a pop-up single-button widget in Section~\ref{sub:validationagent} to, instead, show multiple buttons where each button navigates to its respective activity by invoking the respective instrumentation script (Step \numblack{14}) as depicted in Figure~\ref{fig:instrumentedexploration}. 
To remain as faithful to real-world usage scenarios as possible, the transitions aided by the instrumented edges must happen only between two activities with a pre-existing caller-callee relationship according to the CTG of the app, which was extracted in Section~\ref{sub:features}.
This way, we further differentiate the dynamic widget instrumentation module from the button widget injection of the script validation agent by injecting activities other than the main activity that has the potential to transition to the respective target activities.
In addition to the buttons that transition to the target activities, the dynamic widget instrumentation module also includes a ``Cancel'' button in the pop-up widget to give the fuzzer an option to dismiss the pop-up widget and continue exploring the features of the current activity.

Algorithm~\ref{alg:widgetinstrumentation} shows how the dynamic widget instrumentation module uses the CTG to find the source activity for each target activity to instrument the button to launch the target Activity on the pop-up button widget that shows the source activity.
The $activity\_dialog$ map in line 2 stores source activities and the set of target activities that are supposed to be launched from the instrumented button shown at the start of the source activity.
In $non\_main\_reachables$ in line 4 is the set of activities that are initially reachable by the GUI fuzzer, and is the key for preserving realistic activity transitions. In lines 13-15, if the source Activity of a target belongs to $non\_main\_reachable$, the source is instrumented with the button that navigates to the target activity.
If there is no source activity found, in the cases of alternate entries, for example, in lines 8-11 and 16-18, main activities of the app are instrumented with the button instead to still ensure reachability of those target activities.

\subsubsection{Concurrent Automated GUI Exploration}
\label{sub:finalexploration}

Once the dynamic widget instrumentation module has injected the edge instrumentation scripts into the app process, \nametool runs a GUI fuzzer to collect coverage enhanced by the instrumentation scripts (Step \numblack{15}). 
\nametool complements the existing GUI fuzzer, which runs in a separate process from dynamic widget instrumentation module that uses the Frida server to inject the instrumentation scripts.
During exploration, the fuzzer interacts with injected button widgets, allowing it to navigate and explore their respective previously unreachable activities. 
This is because each of the injected button widget provides an entry point that bypasses the unreachability constraints to the target activity.


\section{Evaluation}
\label{sec:results}

\subsection{\nametool Implementation and Experimental Setup}
\label{sec:implement}

\subsubsection{\nametool}

For the reasoning engines inside both the static analysis agent and the dynamic edge instrumentation agent, we used Anthropic's API-based Claude Sonnet 3.7 LLM. This model was selected because at the time of our experiments, it demonstrated the best agentic coding performance, and reliably executing tool-calling functions. Our choice is further motivated by Anthropic's foundational work in proposing Model Context Protocol (MCP) that enables such advanced model-tool interactions~\cite{mcpprotocol}. Hence, we also rely on Anthropic's MCP to enable tool-calling support for the agents to access their semantic and episodic memories.

\subsubsection{Baselines}
For RQ1's baseline method, we use APE~\cite{ape} that is used by Akinotcho et al.~\cite{akinotcho2025curse} to empirically establish the 30\% ceiling. APE is selected by Akinotcho et al. for being a performant and widely adopted tool for GUI fuzzing and exploration~\cite{su2021benchmarking}.

For RQ2's baseline, we use LLMDroid is the state-of-the-art for LLM-enhanced GUI-fuzzer. It extends pre-existing tools such as Droidbot, Humanoid and Fastbot by modifying those tools to integrate LLM query during the app exploration.
We choose LLMDroid-Fastbot combination as it provides better coverage compared to other combinations such as LLMDroid-Droidbot and LLMDroid-Humanoid.
Unlike LLMDroid that requires the modification of existing tool, \nametool complements any given existing tool as a separate concurrent process, and does not require to modify any aspect of any tool to enhance their coverage performance.
As such, we are able to evaluate \nametool by complementing Fastbot which LLMDroid is built on top of for a direct comparison with LLMDroid in terms of enhancing coverage performance of existing tools. 

For RQ3, to  evaluate the activity launching effectiveness of instrumentation scripts generated by the dynamic instrumentation agent of \nametool in insolation, Scenedroid is selected for being the state-of-the-art approach in activity launching.

For RQ4, since we are the first automated approach for inferring activation conditions of activities there are no existing automated tools to be used as baseline. Hence, we include a random baseline as a sanity check to ensure \nametool is detecting patterns instead of making random choices.

\subsubsection{Dataset}
As we aim to investigate whether \nametool breaks the 30\% activity coverage ceiling reported by Akinotcho et al. ~\cite{akinotcho2025curse}, we use the same benchmark apps used in Akinotcho et al. ~\cite{akinotcho2025curse}. 
Akinotcho et al. selected 11 apps from most popular list on Google Play Store, and F-droid, and manually labeled the reasons for unreachable activities by GUI fuzzers. These apps are selected for their "diversity and representativeness" in terms of the types of unreachable activities.

For a more comprehensive experiment, we additionally use the apps from AndroTest benchmark~\cite{choudhary2015automated}. AndroTest apps were statically instrumented by the benchmark's original authors to track class, method and line coverage, using the Emma coverage measurement tool~\cite{rubtsov2006emma}. Therefore, for AndroTest benchmark apps, we can also compare class, method and line coverages in addition to activity coverage.
The AndroTest benchmark contains 66 apps but we filtered 16 apps for four reasons: (1) for 33 apps, the 100\% activity coverage was already achieved by the existing tools, Ape and Fastbot; (2) 4 apps have newer versions present in either Akinotcho et al.'s selected apps, or AndroTest benchmark; (3) for 4 apps, their CTG - an essential input for \nametool - cannot be extracted; (4) for 9 apps, the Dynamic Edge Instrumentation agent is not able to generate valid script within 5 attempts; while we acknowledge that this could be a limitation of our underlying tool, the 9 apps have an average of 3.6 activities. Thus, activity coverage is not a concern for such small apps.

We refer to the dataset of $27$ apps that combines Androtest apps and Akinotcho et al's selected apps as $Dataset_{All}$, and the dataset with only the apps selected by Akinotcho et al. as $Dataset_{Akinotcho}$.

\subsubsection{Reproducing coverage}Some of the apps in Akinotcho et al.'s study~\cite{akinotcho2025curse} are region-specific (McDonald’s Canada, LocalNews Canada, and Loblaw’s PCHealth). Moreover, there is time difference between their experiment and ours during which server-side changes could affect the accessibility of some activities. Thus, to faithfully compare coverage, we re-ran the baseline approaches in the same environment as \nametool.

\subsubsection{Experimental Setup}

The experiments were conducted on an Ubuntu 24.04.1 LTS server equipped with 256 GB of RAM and AMD 48-Core processor running emulated Android version 9 devices.
Following previous works~\cite{mao2016sapienz,akinotcho2025curse,choi2013guided,su2021benchmarking,lai2019goal}, we run all the fuzzers for 60 minutes each.

\label{sub:rq1}
\begin{table*}[t]
\centering
\captionsetup{justification=centering}
\caption{Test Coverage Results Comparison for activity, class, method, and line metrics:\\\nametool-APE (CA) vs APE (A) Approach}
\label{tab:coverage_comparison}
\resizebox{\linewidth}{!}{%
\scriptsize
\renewcommand{\arraystretch}{1.2} 
\begin{tabular}{|p{3cm}|c|cc|cc|cc|cc|}
\hline
\multirow{3}{*}{\textbf{Package Name}} & \multirow{3}{*}{\textbf{\# Act}} & \multicolumn{2}{c|}{\textbf{Activity (\%)}} & \multicolumn{2}{c|}{\textbf{Class (\%)}} & \multicolumn{2}{c|}{\textbf{Method (\%)}} & \multicolumn{2}{c|}{\textbf{Line (\%)}} \\
\cline{3-10}
 
 &  & \textbf{CA} & \textbf{A} & \textbf{CA} & \textbf{A} & \textbf{CA} & \textbf{A} & \textbf{CA} & \textbf{A} \\
 
\hline
\hline
addi & 4 & \textbf{75} & 50 &\textbf{16} & 15 & \textbf{15} & 12 & \textbf{22} & 19 \\
\hline
keepass & 14 & \textbf{64} & 36 & \textbf{27} & 20 & \textbf{14} & 10 & \textbf{10} & 7 \\
\hline
alarmclock & 5 & \textbf{100} & 80 & \textbf{90} & 16 & \textbf{80} & 10 & \textbf{72} & 15 \\
\hline
bequick & 11 & \textbf{100} & 91 & \textbf{65} & 64 & \textbf{54} & 54 & \textbf{55} & 53 \\
\hline
mileage & 50 & \textbf{94} & 74 & \textbf{84} & 70 & \textbf{75} & 61 & \textbf{63} & 51 \\
\hline
photostream & 4 & \textbf{100} & 50 & \textbf{51} & 39 & \textbf{40} & 27 & \textbf{35} & 25 \\
\hline
acal & 26 & \textbf{77} & 73 & \textbf{40} & 39 & \textbf{36} & 34 & \textbf{28} & 27 \\
\hline
syncmypix & 7 & \textbf{86} & 71 & \textbf{67} & 57 & \textbf{41} & 30 & \textbf{29} & 21 \\
\hline
jamendo & 14 & \textbf{71} & 64 & \textbf{55} & 53 & \textbf{36} & 34 & \textbf{29} & 28 \\
\hline
whohasmystuff & 3 & \textbf{100} & 67 & \textbf{89} & 74 & \textbf{81} & 59 & \textbf{69} & 45 \\
\hline
adsdroid & 2 & \textbf{100} & 50 & \textbf{75} & 50 & \textbf{60} & 38 & \textbf{43} & 28 \\
\hline
myLock & 4 & \textbf{75} & 25 & \textbf{60} & 42 & \textbf{43} & 29 & \textbf{40} & 29 \\
\hline
LNM & 4 & \textbf{75} & 50 & \textbf{57} & 39 & \textbf{47} & 35 & \textbf{44} & 31 \\
\hline
netcounter & 3 & \textbf{100} & 33 & \textbf{51} & 27 & \textbf{39} & 22 & \textbf{33} & 17 \\
\hline
fantastischmemo & 34 & \textbf{74} & 68 & \textbf{59} & 56 & \textbf{51} & 48 & \textbf{48} & 45 \\
\hline
aagtl & 3 & \textbf{100} & 33 & \textbf{19} & 16 & \textbf{11} & 10 & \textbf{17} & 15 \\
\hline
\textbf{Average Coverage} &  & \textbf{86.9} & 60.3 & \textbf{56.6} & 42.3 & \textbf{45.2} & 32.1 & \textbf{39.8} & 28.5 \\
\hline
\hline
wordpress & 111 & \textbf{47} & 3 & - & - & - & - & - & - \\
\hline
pchealth & 29 & \textbf{48} & 3 & - & - & - & - & - & - \\
\hline
mcdonalds & 137 & \textbf{25} & 9 & - & - & - & - & - & - \\
\hline
pinterest & 30 & \textbf{37} & 10 & - & - & - & - & - & - \\
\hline
pluto & 13 & \textbf{77} & 77 & - & - & - & - & - & - \\
\hline
easyMover & 51 & \textbf{49} & 27 & - & - & - & - & - & - \\
\hline
myexpenses & 51 & \textbf{59} & 41 & - & - & - & - & - & - \\
\hline
localnews & 75 & \textbf{61} & 5 & - & - & - & - & - & - \\
\hline
temu & 33 & \textbf{42} & 15 & - & - & - & - & - & - \\
\hline
wikipedia & 48 & 67 & 67 & - & - & - & - & - & - \\
\hline
k9 & 33 & \textbf{33} & 6 & - & - & - & - & - & - \\
\hline
\textbf{Average Coverage} &  & \textbf{49.5} & 17.7 &  &  &  &  &  &  \\

\hline
\hline
\textbf{Combined Avg. Cov.} &  & \textbf{71.7} & 41.1 & \textbf{56.6} & 42.3 & \textbf{45.2} & 32.1 & \textbf{39.8} & 28.5 \\
\hline
\end{tabular}
}
\end{table*}

\subsection{RQ1: Breaking the Ceiling}
\label{sub:rq1}

Our first research question is regarding whether \nametool achieves a breakthrough in activity coverage by breaking the ceiling of 30\% that was presented by Akinotcho et al. \cite{akinotcho2025curse}. We pick APE, the open-source and widely used GUI exploration tool that was used in automated exploration of applications by Akinotcho et al. to establish the 30\% coverage ceiling. 
First, we run APE on $Dataset_{All}$. We then \nametool's instrumented exploration on the same dataset with APE as the complementing exploration tool.
To prove that \nametool is designed to enhance GUI-fuzzers to achieve higher coverage across all metrics, we also use class, method and line coverage percentages which measures the proportion of the app classes, methods and lines of code were executed during testing. These metrics, together with activity coverage, collectively provide a comprehensive overview of how well the app under test is exercised from high-level UI interaction to low-level code executions.

Table~\ref{tab:coverage_comparison} presents the comparative analysis result of test coverage between the baseline APE and \ourapproach-APE. The results show that APE when complemented with \ourapproach consistently outperforms the baseline, i.e., APE alone, across all four metrics, thus breaking the 30\% activity coverage ceiling for GUI-fuzzers. 
More precisely, for apps in $Dataset_{Akinotcho}$ that is a popular representative of everyday real-world apps, our tool achieves 49.5\% activity coverage compared to 17.7\% achieved by APE, which is a $179.7\%$ improvement.
This highlights the strong capabilities of \ourapproach in enhancing existing tools to improve their activity coverage coverage.

In addition, \nametool outperforms APE in class (56.6\% vs 42.3\%), method (45.2\% vs 32.1\%), and line (39.8\% vs 28.5\%) average coverage. This shows that \nametool not only achieves breakthrough in activity coverage, but also ensures the lower-level code coverage for more thorough testing of the app. \nametool achieves this by outperforming APE not only in average scores, but also in each of the individual apps. For example, \ourapproach-APE achieves 100\% coverage in several of the apps where APE alone achieves significantly lower activity coverage. Even in complex apps with several activities, \ourapproach consistently improves APE by 20-40 percentage points (180\%-1050\% improvements).

\begin{tcolorbox}
{\bf Answer to RQ1:} 
\nametool-APE achieves a significant improvement in terms of the activity coverage over APE alone, achieving an average of 49.5\% activity coverage compared to 17.7\% by APE. Furthermore, \nametool-APE outperforms APE on other coverage metrics, with \nametool-APE achieving 33.7\% class, 40.8\% method, and 39.6\% line coverage improvements over APE.
\end{tcolorbox}

\subsection{RQ2: Comparison Against the State-of-the-Art} 

To evaluate the performance of \nametool against the state-of-the-art approach, namely LLMDroid, we use the same dataset, $Dataset_{All}$ and same metrics as {\bf RQ1}.
Since, LLMDroid is implemented by extending existing tools (Droidbot, Humanoid, and Fastbot), we picked Fastbot which is the tool that LLMDroid performed the best with. 
As another baseline and also as a sanity check, we run Fastbot alone using apps from $Dataset_{All}$. We then run LLMDroid-Fastbot and \nametool-Fastbot independently on the benchmark apps.

Table~\ref{tab:llmdroidfastbotcomparison} shows the comparison between \nametool-Fastbot, LLMDroid-Fastbot, and the baseline Fastbot in terms of activity, class, method and line coverage. 
\nametool-Fastbot is able to outperform, on average, both LLMDroid-Fastbot and Fastbot in all metrics. 
In particular, on apps from $Dataset_{Akinotcho}$, \nametool-Fastbot delivers $116.3\%$ higher activity coverage (34.6\% vs 17.2\%) compared to LLMDroid-Fastbot. For example, for complex apps with multiple Activities such as \emph{mileage} and \emph{acal}, \nametool-Fastbot achieves substantially higher activity coverage subsequently leading to a higher class, method, and line coverage.
Likewise for more granular code coverage metrics, \nametool-Fastbot outperforms both LLMDroid-Fastbot and Fastbot in average coverage score by achieving 27.3\% higher for class coverage, 31.9\% higher for method coverage and 29.2\% higher line coverage.

\begin{table*}[t]
\centering
\captionsetup{justification=centering}
\caption{Test Coverage Results Comparison for activity, class, method, and line metrics: \\ \nametool-Fastbot (CF) vs LLMDroid-Fastbot (LF) vs Fastbot (F) Approach}
\label{tab:llmdroidfastbotcomparison}
\resizebox{\linewidth}{!}{%
\small
\renewcommand{\arraystretch}{1.2} 
\begin{tabular}{|p{2.5cm}|c|ccc|ccc|ccc|ccc|}
\hline
\multirow{3}{*}{\textbf{Package Name}} & \multirow{3}{*}{\textbf{\# Act}} & \multicolumn{3}{c|}{\textbf{Activity (\%)}} & \multicolumn{3}{c|}{\textbf{Class (\%)}} & \multicolumn{3}{c|}{\textbf{Method (\%)}} & \multicolumn{3}{c|}{\textbf{Line (\%)}} \\
\cline{3-14}
 &  & \textbf{CF} & \textbf{LF} & \textbf{F} & \textbf{CF} & \textbf{LF} & \textbf{F} & \textbf{CF} & \textbf{LF} & \textbf{F} & \textbf{CF} & \textbf{LF} & \textbf{F} \\
\hline
\hline
addi & 4 & 25 & 25 & 25 & 12 & 12 & 12 & 10 & 10 & 10 & \textbf{18} & 17 & 17 \\
\hline
keepass & 14 & \textbf{43} & 28 & 14 & \textbf{22} & 16 & 16 & \textbf{11} & 8 & 2 & \textbf{8} & 5 & 1 \\
\hline
alarmclock & 5 & 80 & 80 & 60 & \textbf{86} & 84 & 84 & \textbf{77} & 74 & 54 & \textbf{69} & 66 & 48 \\
\hline
bequick & 11 & \textbf{91} & 63 & 72 & \textbf{57} & 49 & 49 & \textbf{51} & 39 & 42 & \textbf{51} & 39 & 41 \\
\hline
mileage & 50 & \textbf{50} & 20 & 16 & \textbf{65} & 40 & 40 & \textbf{44} & 34 & 32 & \textbf{44} & 30 & 27 \\
\hline
photostream & 4 & 50 & 50 & 25 & \textbf{37} & 16 & 16 & \textbf{26} & 11 & 2 & \textbf{23} & 13 & 1 \\
\hline
acal & 26 & \textbf{73} & 27 & 15 & \textbf{43} & 23 & 23 & \textbf{38} & 15 & 12 & \textbf{30} & 11 & 9 \\
\hline
syncmypix & 7 & \textbf{86} & 62 & 71 & \textbf{67} & 55 & 55 & \textbf{42} & 29 & 26 & \textbf{31} & 20 & 18 \\
\hline
jamendo & 14 & \textbf{57} & 43 & 36 & \textbf{45} & 40 & 40 & \textbf{32} & 26 & 21 & \textbf{26} & 20 & 15 \\
\hline
whohasmystuff & 3 & 100 & 100 & 100 & 89 & \textbf{93} & 93 & 78 & 78 & 75 & \textbf{68} & 65 & 65 \\
\hline
adsdroid & 2 & \textbf{100} & 50 & 50 & \textbf{75} & 33 & 33 & \textbf{60} & 28 & 15 & \textbf{42} & 20 & 8 \\
\hline
myLock & 4 & \textbf{50} & 25 & 25 & \textbf{47} & 42 & 42 & \textbf{33} & 26 & 25 & \textbf{32} & 28 & 26 \\
\hline
LNM & 4 & \textbf{75} & 50 & 50 & \textbf{57} & 39 & 39 & \textbf{48} & 34 & 29 & \textbf{44} & 30 & 23 \\
\hline
netcounter & 3 & \textbf{100} & 33 & 67 & \textbf{46} & 28 & 28 & \textbf{35} & 23 & 23 & \textbf{30} & 18 & 18 \\
\hline
fantastischmemo & 34 & 50 & \textbf{62} & 24 & 39 & \textbf{48} & 48 & 34 & \textbf{42} & 22 & 31 & \textbf{39} & 18 \\
\hline
aagtl & 3 & \textbf{67} & 33 & 33 & \textbf{19} & 16 & 16 & \textbf{12} & 10 & 10 & \textbf{20} & 17 & 17 \\
\hline
\textbf{Average Coverage} &  & \textbf{69.1} & 47.6 & 38.1 & \textbf{50.4} & 39.6 & 32.9 & \textbf{40.1} & 30.4 & 25.0 & \textbf{35.4} & 27.4 & 22.0 \\
\hline
\hline
wordpress & 111 & \textbf{10} & 3 & 3 & - & - & - & - & - & - & - & - & - \\
\hline
pchealth & 29 & \textbf{38} & 4 & 4 & - & - & - & - & - & - & - & - & - \\
\hline
mcdonalds & 137 & \textbf{22} & 4 & 4 & - & - & - & - & - & - & - & - & - \\
\hline
pinterest & 30 & \textbf{23} & 13 & 10 & - & - & - & - & - & - & - & - & - \\
\hline
pluto & 13 & \textbf{54} & 8 & 8 & - & - & - & - & - & - & - & - & - \\
\hline
easyMover & 51 & \textbf{35} & 20 & 22 & - & - & - & - & - & - & - & - & - \\
\hline
myexpenses & 51 & \textbf{51} & 28 & 20 & - & - & - & - & - & - & - & - & - \\
\hline
localnews & 75 & \textbf{44} & 37 & 37 & - & - & - & - & - & - & - & - & - \\
\hline
temu & 33 & \textbf{27} & 15 & 15 & - & - & - & - & - & - & - & - & - \\
\hline
wikipedia & 48 & \textbf{67} & 52 & 48 & - & - & - & - & - & - & - & - & - \\
\hline
k9 & 33 & \textbf{9} & 6 & 6 & - & - & - & - & - & - & - & - & - \\
\hline
\hline
\textbf{Average Coverage} &  & \textbf{34.6} & 17.2 & 16.0 & - & - & - & - & - & - & - & - & - \\
\hline
\hline
\textbf{Comb. Avg. Cov.} &  & \textbf{55.1} & 35.2 & 29.1 & \textbf{50.4} & 39.6 & 32.9 & \textbf{40.1} & 30.4 & 25.0 & \textbf{35.4} & 27.4 & 22.0\\
\hline
\end{tabular}

}
\end{table*}

\subsubsection{Case Studies}\leavevmode\\

\begin{figure*}[h]
    \includegraphics[width=1\textwidth, trim=0cm 0.25cm 0cm 0cm]{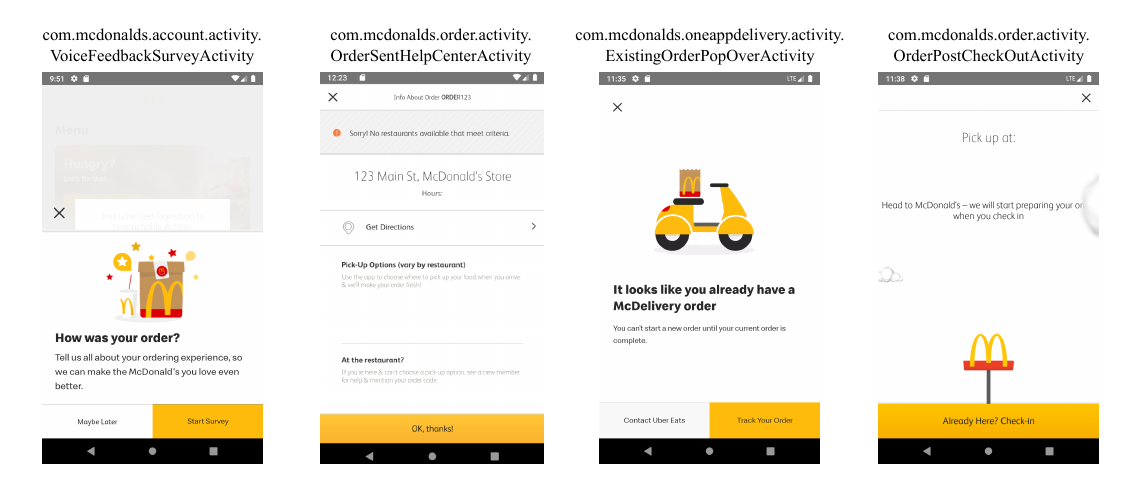}
    \vspace{-1em}
\captionsetup{justification=centering}
    \caption{Example activities in McDonald's Canada app that are unreachable by existing approaches (LLMDroid, Ape, and Fastbot), but reachable by \nametool's dynamically instrumented edges}
    \label{fig:rq1case2}
\end{figure*}

\textbf{\texttt{mcdonalds}.} We take a case study when exploring the McDonald's Canada app exemplifies the advantages of edge instrumentations in GUI exploration. 
The newly explored target activities after running \nametool-Fastbot, \texttt{OrderSentMapActivity}, and \texttt{OrderPostCheckoutActivity} activities (shown in Figure~\ref{fig:rq1case2}), are only reachable after making the order and payments from McDonald's. 
In order to reach this activity under normal circumstances, the human tester or the UI fuzzer have to make an order by making actual payments and selecting delivery destinations, and receive respond from the server that allows the transition to \texttt{OrderSentMapActivity}, or \texttt{OrderPostCheckoutActivity}.
LLMDroid, although guided by an LLM capable of reasoning with complex UI screens, was not able to navigate through a series of other activities leading to the target activities reachable by \nametool.
\nametool is able to resolve all the activation conditions and data dependencies in terms of order and payment status from the previous activity, to facility a successful transition to those activity by an concurrently running existing UI fuzzer tool.

\textbf{\texttt{fantastischmemo}.} As discussed in Section~\ref{sub:instrumetation}, 
\nametool only injects the button on a non-main source activity that has outgoing edge to the unreachable target activity according to the CTG.
\emph{fantastischmemo} is an old-fashioned app with Tab activity that somehow only invokes the onResume hook when the Tab activity is resumed from "Recents" screen.
However, the injected button widgets are not shown when navigating directly to the Tab activity from other Tab activities.
This can be attributed to our widget instrumentation script not being complete/robust enough to handle apps with legacy APIs such as TabActivities.
However, this is not a big enough issue in terms of generalizability because the TabActivity API has been deprecated since Android 3.

\textbf{\texttt{addi}} and \textbf{\texttt{photostream}.} During the exploration of \emph{addi} and \emph{photostream} app, the injected pop-up widget is shown numerous times to enable transitions to the target activities. 
However, every time the widget is shown, the Fastbot tool chooses to click on "Cancel" to dismiss the widget, and continue exploring the activities that have already been discovered. 
Thus, our framework is unable to reach new activities that are not yet reached by Fastbot, since Fastbot itself chooses to not click on the other buttons to transition to the respective target activities.
This is in contrast to APE tool which chooses not to cancel the pop-up and, instead, uses the provided buttons to improve its exploration coverage. 
The exact reason behind the difference in exploration behaviors between APE and Fastbot can be attributed to difference in their implementations, and is out of the scope of this work.

\begin{tcolorbox}
{\bf Answer to RQ2:} \nametool-Fastbot significantly improves test coverage over the state-of-the-art tool, LLMDroid-Fastbot, achieving average of 50.1\%, 26.4\%, 35.6\%, and 31.6\% improvements for activity, class, method and line coverage respectively. This result highlights the effectiveness of \nametool in enhancing GUI-fuzzers compared to the state-of-the-art approach.

\end{tcolorbox}

\subsection{RQ3: Activity Launching}
\label{sub:rq3}

We aim to measure the performance of LLM, which is at the core of our approach in the context of generating Frida instrumentation scripts to instrument Android apps. 
More specifically, we focus on the activity launching topic of Android app analysis, and compare the dynamic instrumentation agent of \nametool against the state-of-the-art approach, Scenedroid~\cite{scenedroid} in that topic. 
Scenedroid uses ICC information extracted by ICCBot~\cite{yan2022iccbot} to prepare ICC messages and launch activities via ADB.
As such, the metric for comparison is the activity launching success rate.
We perform this comparison on the manually-labelled activities of apps from $Dataset_{Akinotcho}$.
The unreachability of an activity can be attributed to a combination of single or multiple reasons that are manually-labelled by Akinotcho et al. 
We refer to such combinations as \textit{categories}, we can compare activity launching capability of our instrumented edges to that of Scenedroid. 

\begin{figure}[t]
    \includegraphics[width=1\textwidth, trim=0.7cm 0.5cm 0.5cm 0.15cm]{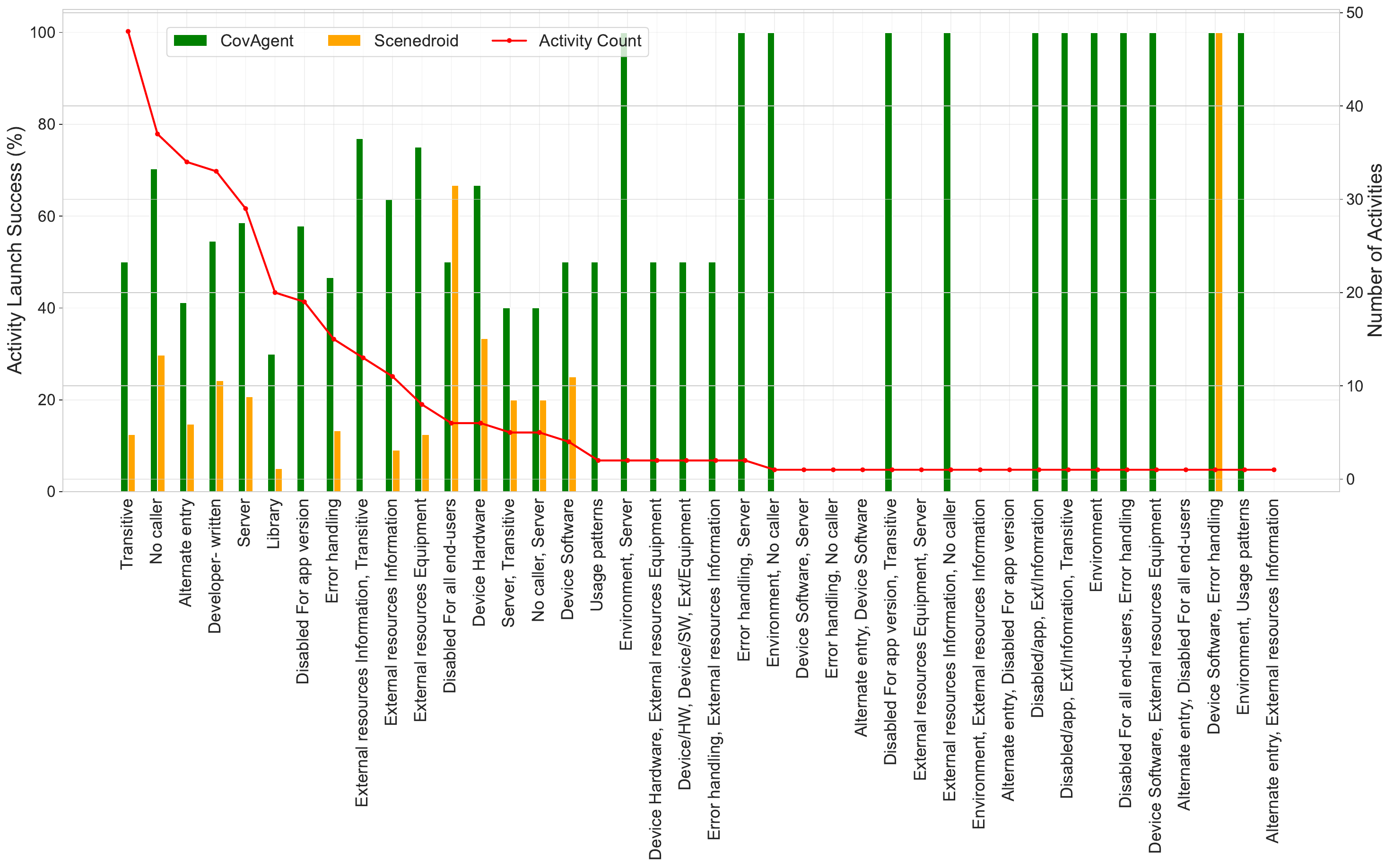}
    \caption{Activity Launching Success Rate}
    \label{fig:RQ3}
\end{figure}

Figure~\ref{fig:RQ3} presents a bar chart that shows the comparison of activity launch success rate between \nametool and Scenedroid. It also contains line graph that shows the number of activities for each category with the categories sorted along the x-axis according to their number of activities. Thus, 18 categories on the right gives a numerically unstable score of $0$ or $1$, but we still include them in this chart for the sake of completeness. Conversely, the categories on the left have more numerically stable activity launch success rate for both \nametool and Scenedroid. 

In all categories that contain at least 2 activities, \nametool outperforms Scenedroid in activity launching except for the "Disabled for all end-users" category. More specifically, \nametool achieves 54.8\% launch success in weighted average across all categories, compared to Scenedroid's weighted average of 15.8\%. 
In particular, the "External resources (Information), Transitive" category, \nametool have nearly $80\%$ activity launch success rate, while Scenedroid fails absolutely at $0\%$. 
Upon further investigation, the activities are all related to the ordering process of McDonald's Canada app that requires location information of the user which maps to the "External resources (Information)" aspect of the category. Scenedroid fails because this location information is not resolved from ICC message that is sent by source activity, but rather by invoking the \texttt{RestaurantDependencyWrapper.a()} API. 
Scendroid can only prepare the contents of Intent messages sent to the target activity, but not modify (instrument) the perception inside the target activity. 
Meanwhile, this "external resource" dependency can be resolved by \nametool where the instrumentations hook the \texttt{RestaurantDependencyWrapper.a()}'s implementation to return mock location coordinates, allowing the component to proceed with its initialization procedures and launch the target activity successfully. 
\textbf{Metaphorically speaking, while Scenedroid attempts to craft the key (Intent message) to unlock the door (target activity), \nametool removes the lock itself (blocking guard conditions) via dynamic instrumentation.}

For the "Disabled for all end-users" category, $2$ out of $6$ activities are launchable by Scenedroid but not by \nametool, and the remaining 4 are launchable by both approaches. Those are Wordpress' \texttt{DebugSettingsActivity} and Temu's \texttt{PermissionRequestActivity}. In both cases, we found that the dynamic instrumentation agent received insufficient activation conditions received from the static analysis agent in the prior phase.
Even though the inferred activation conditions mention the accurate unreachability reason, "Disabled for end-users", it could not provide accurate guidance how to generate instrumentations. This can be attributed to dynamic edge instrumentation agent's reliance on the output of static analysis agent which is also LLM-driven and has to be accounted for its stochasticity.

\begin{tcolorbox}
{\bf Answer to RQ3:} \nametool's activity launching capability through its dynamic edge instrumentation agent achieves 54.8\% activity launch success rate which vastly outperforms the state-of-the-art Scenedroid's 15.8\%. 
This establishes LLM-driven AI agents' role in using its code generation capabilities to bypass roadblocks in the topic of Android testing.
\end{tcolorbox}

\subsection{RQ4: Activity Activation Condition Inference}
\label{sub:rq4}

To evaluate the performance of \nametool's static analysis agent in inferring the activity activation conditions, we need to compare the inferred activation conditions against a ground truth. The ground truth can be the manually-labelled activities by Akinotcho et al. \cite{akinotcho2025curse} from the $Dataset_{Akinotcho}$.
We formulate this as a ranking problem where recall is a common evaluation metric to measure how much of the ground truth is identified for a set of chosen top-k predictions.
As such, we use recall as the evaluation metric measuring the percentage of the ground truth unreachability reasons in a manually-labelled activity that is correctly identified by \nametool.

To convert the natural language format of the inferred activation conditions to unreachability reasons that is directly comparable with ground truth, 
we use an LLM to predict a top-5 closest reasons out of the 13 unreachability reasons based a given activation condition description.
We include the description of each of the reasons to guide the LLM with its ranking.
As a sanity check, we compare \nametool against random selection ensuring that \nametool's performance reflects meaningful information rather than a mere chance.

\begin{figure}
    \centering
    \includegraphics[width=\linewidth,trim=0cm 1.8cm 0cm 0.5cm]{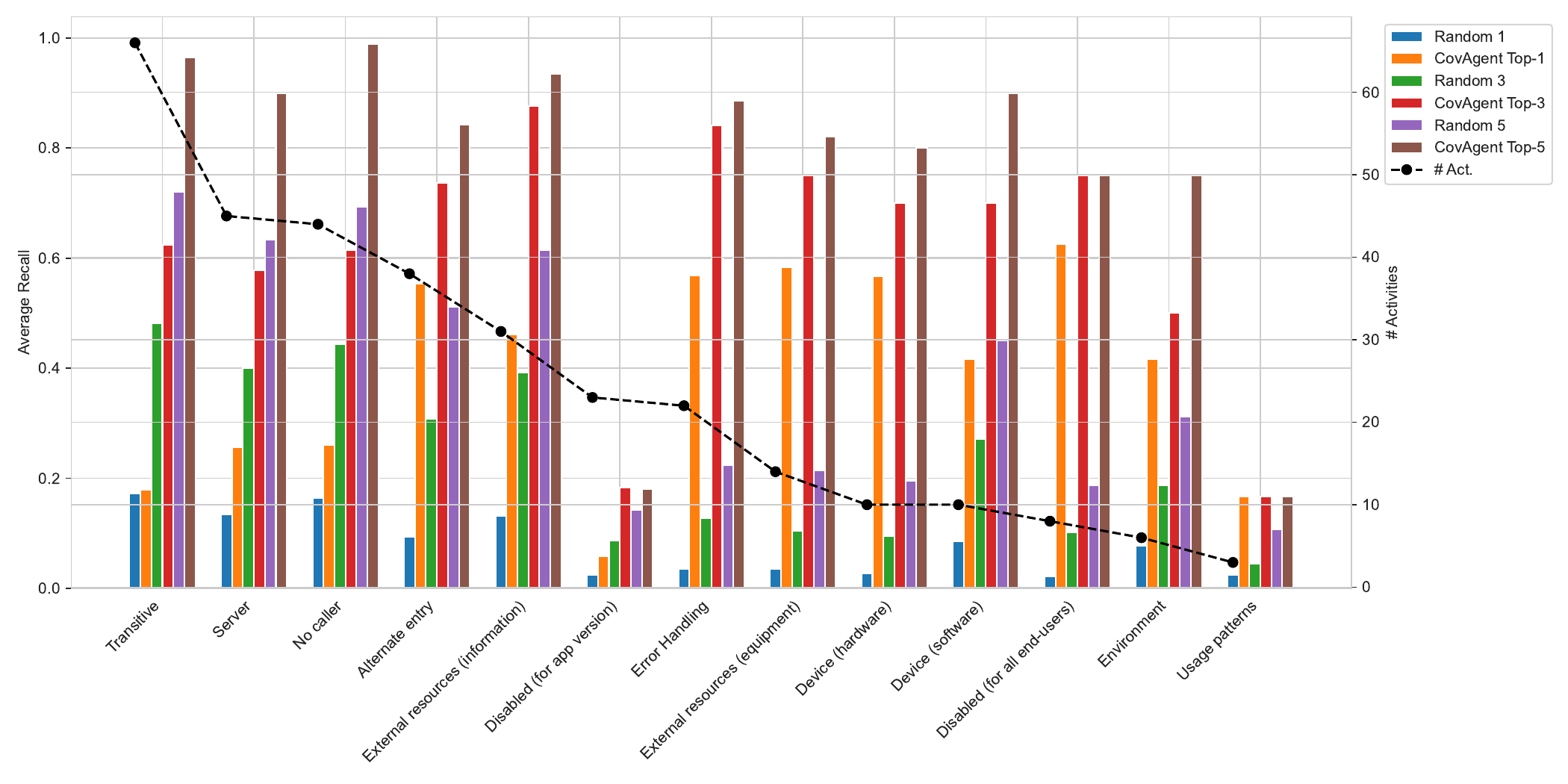}
    \vspace{-1em}
    \caption{Activation condition extraction accuracy}
    \label{fig:rq4}
\end{figure}

Figure~\ref{fig:rq4} compares the accuracy of \nametool's activation condition extraction against the ground truth. The y-axis on the left shows average recall, while the y-axis on the right shows the number of Activities per activation condition category. The results show that \nametool consistently outperforms random baselines in automatically extracting activation conditions. Random selection results in low recall across all categories while \nametool achieves a higher accuracy with Top-1 prediction already surpassing Random-5 in 54\% of the categories. Performance further improves with Top-3 and Top-5 predictions where recall increases from 0.6 to 0.9 in most categories indicating \nametool's capabilities in extracting the correct activation condition among the top candidates. We also observe \nametool's performance in detecting the right category candidates for the most common activity activation conditions. These are the categories with high number of activities such as Transitive,  Server, and No-caller conditions.

\begin{tcolorbox}
{\bf Answer to RQ4:} \nametool significantly performs better than a random baseline in activation condition extraction achieving on average 0.34 (Top-1), 0.64 (Top-3), and 0.85 (Top-5) recall vs 0.11, 0.33, and 0.52, respectively. Compared to the ground truth provided by Akinotcho et. al., \nametool achieves consistently higher recall in the most common activation conditions demonstrating agentic inference can serve as a reliable alternative to labor-intensive manual investigation.
\end{tcolorbox}

\section{Discussion}

\subsection{Limitations}

\subsubsection{Dependence on accompanying fuzzer to interact with GUI} In Section 4.3.1, we presented cases where the Fastbot tool does not always interact with the injected button widgets that transition to the target activities. This is a bottleneck in CovAgent which relies on accompanying fuzzer, Ape~\cite{ape}, and Fastbot~\cite{lv2022fastbot2}, to interact with the widgets and trigger transition to the target activity. 
However, \textbf{our work focuses on satisfying the activation conditions of unreachable activities, and investigating the peculiarities of the accompanying fuzzer such as Fastbot is out of the scope of this work}.
Moreover, this limitation can be overcome by using a more performant accompanying fuzzer, Ape, which is shown to interact with the injected button widget more comprehensively than Fastbot.
\nametool readily and seamlessly integrates with any accompanying fuzzer as it runs in a separate process, thus the possibility to integrate with any new state-of-the-art accompanying fuzzer released in the future as well to keep improving coverage.

\subsubsection{Second-order feasibility of generated parameters values} In the scope of our work, we focus on generating the edge instrumentation scripts satisfy the activation conditions to successfully launch the respective target activities, which would otherwise never be launched by existing approaches such as LLMDroid~\cite{llmdroid}, Ape~\cite{ape}, and Fastbot~\cite{lv2022fastbot2} (as demonstrated in Section~\ref{sec:results}). 
We recognize the importance of the generating parameter values that also comprehensively initialize all the features inside the target activity (e.g. populating dropdown lists, and other views with generated data) to perfectly simulate real-life usage environment after successfully launching the target activity. 
As such, we leveraged the AndroTest dataset that comes with Emma-instrumentation to \textbf{measure finer-grained class, method and statement coverage, and verified \nametool's superior exploration capability compared to the baselines: LLMDroid, Fastbot, and Ape}. 
A more in-depth analysis of LLM-driven agents' capacity for comprehensive initialization of all the features inside the target activity is left for future work.

\subsection{Threats to Validity}

\subsubsection{Internal Validity.} First, like other approaches that leverages LLMs, our approach is also affected by the stochasticity in the output of LLMs. To mitigate this, we design an iterative process to validate the script, and provide feedback containing exception messages from the environment for better-informed refinement of the script.
Secondly, our evaluations in Section~\ref{sub:rq3} and Section~\ref{sub:rq4} require ground-truth labelings of activities for their unreachability reason(s). To ensure the correctness of ground truth, we pick data provided by Akinotcho et al.~\cite{akinotcho2025curse} who recruited an external mobile analyst to manually and systematically analyze the activities.

\subsubsection{External Validity.} We evaluated \nametool on 27 apps from the Androtest dataset, and those selected by Akinotcho et. al~\cite{akinotcho2025curse}. %
The AndroTest dataset was used by prior studies, such as \cite{romdhana2022deep, su2017guided, mao2016sapienz, akinotcho2025curse}, and 
Akinotcho et al.'s dataset was selected from a pool of popular Google Play and F-droid apps to construct a “diverse and representative” sample of apps that comprehensively include all types of unreachable activities.
It is important to note that the number of apps in the dataset is only one aspect of our evaluation (RQ1 and RQ2), since we evaluate RQ3 and RQ4 at the level of individual unreachable (target) activities, not individual apps. 
To this end, there are $363$ unreachable activities that are manually verified and labelled by Akinotcho et. al to ensure that the target activities dataset is diverse and representative.
Thus, the wide adoption of the dataset in the evaluation of previous works, and the representativeness of the apps, and target activities ensured by the Akinotcho et al. lends to the generalizability and representativeness of our dataset.

Another source of concern is regarding the practicality and robustness of \nametool as it relies on Frida, a dynamic instrumentation tool for Android apps. While it is true that Frida can add instability due to apps deploying hardening features and preventing dynamic instrumentation, Frida is still a widely-adopted industrial-strength dynamic instrumentation tool used by prior works for testing and discovering vulnerabilities in Android apps~\cite{otphunter, appprotect}. Thus, our approach is as practical and robust as those prior works.

The obfuscation of apps is a major hinderance to approaches involving the analysis of bytecode (Smali code in our case), and the runtime log messages~\cite{obfuscation}. 
However, we instruct the Static Analysis agent to start its analysis from lifecycle hooks of Android activities such as \texttt{onCreate}, and \texttt{onResume} which cannot be obfuscated as they are part of the Android Framework~\cite{lifecyclehooks}, and not defined by the app developers. 
The runtime exception messages from script validation phase are also associated to the lifecycle hooks as the exceptions are expected at the launch of the target activity.
As such, our approach is able to analyze and improve exploration coverage of obfuscated apps as we only rely on the lifecycle hook APIs that cannot be obfuscated by the app developers.

\subsubsection{Construct Validity.} The effectiveness of \nametool for enhancing Android app testing coverage is adequately evaluated by activity and code coverage metric, and without fault detection metrics. 
We follow a prior work, LLMDroid~\cite{llmdroid}, that justified their exclusion of fault detection metric by highlighting the limitation of fault detection metrics when it comes to evaluating app exploration tools due to their high instability in results even with maximum test rounds used in its previous studies~\cite{testingmetric}.
Hence, fault detection is out of scope of this work, and our future work will extend \nametool’s capabilities towards automated bug detection.

\subsection{Cost}

Since we use Anthropic's Claude Sonnet, a proprietary LLM, to power the agents in our approach, the cost of LLM API invocations is an essential factor when facilitating replicability of our experiment~\footnote{As of the time of our experiment, price per 1M token for Claude Sonnet 3.7 is \$$3.0$/\$$15.0$ (input/output).}. Each activity involves two LLM-driven agents: Static Analysis agent, and Dynamic Edge Instrumentation agent. Static analysis agent spends an average of $77327$/$15205$ (input/output) tokens to infer activation conditions, and costing an average of $0.15$ USD per target activity. Dynamic Edge Instrumentation agent spends an average of $59084$/$35530$ (input/output) tokens, and costs an average of $0.24$ USD per target activity. Since there are total of $842$ target activities across the 27 apps in our dataset, the total cost for all target activities for both agents is $328.45$ USD. 

The high consumption of input and output tokens can be explained by the MCP tool calls made by the agents to inspect the Smali code of the apps. The results of the tool calls goes back to the LLM as input prompt tokens while the LLM generate multiple rounds of output tokens to signal tool invocations to the ADK framework. We acknowledge that the cost incurred for our experiments is not trivial for resource-constrained organizations looking to replicate our study. As such, in our future work, we plan to explore the capability of locally-run lightweight LLM models in AI agents for enabling cost-effective large-scale analysis of bigger Android app datasets.

\section{Conclusion and Future Work}
\label{sec:conclusion}

In this paper, we present \ourapproach, a novel agentic AI-enhanced framework that leverages LLMs to reason about unreachability problem of Android app activities. By employing coordinated static-dynamic analysis, \ourapproach uses LLM-powered agents that use MCP to communicate with different tools in order to reason about why certain activities are unreachable by standard GUI-fuzzers. It then automatically generates dynamic instrumentation scripts to bypass these complex activation conditions such as dependencies on device state or external resources. 
The evaluation results demonstrate that \ourapproach can successfully reason about the unreachability of activities and automatically bypass the complex activation guards, significantly boosting the performance of existing fuzzers such as APE and Fastbot. The results show that this agentic approach is an effective method for improving the reliability of automated mobile app testing. 
Future work will focus on extending \ourapproach's capabilities from coverage enhancement to automated bug detection. By enabling access to previously unreachable paths, \ourapproach could provide unique opportunity to uncover hidden defects.
We also plan to explore the capability of lightweight locally-run LLMs to improve the cost-efficiency of our AI agents, and enable large-scale analysis of a greater number of apps and their target activities.

\section{Data Availability}

Our replication package containing our implementation of \nametool, the datasets we evaluated \nametool on, and the evaluation results are available at \url{https://osf.io/vxgwm/?view_only=3edf2abe5a12419fa6413933da851847}~\cite{replicationpackage}.

\bibliographystyle{ACM-Reference-Format}
\bibliography{main}

\end{document}